\documentclass[fleqn,usenatbib]{mnras}

\usepackage{newtxtext,newtxmath}

\usepackage[T1]{fontenc}

\DeclareRobustCommand{\VAN}[3]{#2}
\let\VANthebibliography\thebibliography
\def\thebibliography{\DeclareRobustCommand{\VAN}[3]{##3}\VANthebibliography}

\usepackage{graphicx}	
\graphicspath{{../}{images/}}	

\usepackage{amsmath}	
\usepackage{amssymb}	
\usepackage[acronym]{glossaries}	

\usepackage{multirow}
\usepackage[table,xcdraw]{xcolor}
\usepackage{soul} 
\usepackage[normalem]{ulem} 

\newcommand{\tprzero}{\,TPR$_0$}	
\newcommand{\tprten}{\,TPR$_{10}$}	

\newacronym{CNN}{CNN}{Convolutional Neural Network}
\newacronym{API}{API}{Application Programming Interface}
\newacronym{KiDS}{KiDS}{Kilo-Degree Survey}
\newacronym{SVM}{SVM}{Support-Vector Machine}
\newacronym{HOG}{HOG}{Histogram of Oriented Gradients}
\newacronym{ReLU}{ReLU}{Rectified Linear Unit}
\newacronym{ELU}{ELU}{Exponential Linear Unit}
\newacronym{TPR}{TPR}{True Positive Rate}
\newacronym{FPR}{FPR}{False Positive Rate}
\newacronym{ROC}{ROC}{Receiver Operating Characteristic}
\newacronym{AUC}{AUC}{Area Under the Curve}
\newacronym{LEXACTUM}{LEXACTUM}{Lens EXtrActor CaTania University of Malta}
\newacronym{OACT}{OACT}{Osservatorio Astrofisico di Catania}
\newacronym{INAF}{INAF}{Istituto Nazionale di Astrofisica}

\defcitealias{zscale}{NOAO 1997}

\title[LEXACTUM]{A Comparative Study of Convolutional Neural Networks for the Detection of Strong Gravitational Lensing}

\author[D. Magro et al.]{
Daniel Magro,$^{1,2}$\thanks{E-mail: daniel.magro.15@um.edu.mt (UM)}
Kristian Zarb Adami,$^{1,2,3}$
Andrea DeMarco,$^{1,2}$
\newauthor
Simone Riggi,$^{2}$
Eva Sciacca$^{2}$
\\
$^{1}$Institute of Space Sciences and Astronomy, University of Malta, Msida MSD2080, Malta\\
$^{2}$Istituto Nazionale di Astrofisica, Osservatorio Astrofisico di Catania, Via S.Sofia 78, Catania 95123, Italy\\
$^{3}$Department of Astrophysics, University of Oxford, Oxford OX1 2JD, United Kingdom
}

\date{Accepted XXX. Received YYY; in original form ZZZ}

\pubyear{2021}

\begin{document}
\label{firstpage}
\pagerange{\pageref{firstpage}--\pageref{lastpage}}
\maketitle

\begin{abstract}
As we enter the era of large-scale imaging surveys with the up-coming telescopes such as LSST and SKA, it is envisaged that the number of known strong gravitational lensing systems will increase dramatically. However, these events are still very rare and require the efficient processing of millions of images. In order to tackle this image processing problem, we present Machine Learning techniques and apply them to the Gravitational Lens Finding Challenge. The \glspl{CNN} presented here have been re-implemented within a new, modular, and extendable framework, LEXACTUM. We report an \gls{AUC} of 0.9343 and 0.9870, and an execution time of 0.0061s and 0.0594s per image, for the Space and Ground datasets respectively, showing that the results obtained by \glspl{CNN} are very competitive with conventional methods (such as visual inspection and arc finders) for detecting gravitational lenses.
\end{abstract}




\section{Introduction}

Strong gravitational lensed systems are unique systems in which a background galaxy and a foreground galaxy or cluster of galaxies are sufficiently well-aligned so that the gravitational field of the foreground system lenses the background galaxies. Whilst these lensing systems hold a rich source of information of the gravitational field distribution of the foreground system and can be used to map dark matter distribution within the cluster, they are rare to come by.
As a matter of fact, \citet{castles_survey} states that the number of known gravitational lenses was 47. The `CASTLES Survey' website\footnote{\url{https://www.cfa.harvard.edu/castles/}}, at the time of writing, lists 100 Multiply Imaged Systems, 92 of which \citet{castles_survey} claim they are confident are lenses. Furthermre, the CLASS \citep{CLASS}, SLACS \citep{SLACS}, H-ATLAS \citep{H-ATLAS}, and SL2S \citep{SL2S} surveys have also contributed to the discovery of gravitational lenses.

Traditional methods for detecting these strongly lensed systems were based on visual inspection and this paper aims to address the automation of this detection. With experiments such as the SKA\footnote{\url{https://www.skatelescope.org/}} \citep{SKA}, LSST \citep{LSST}, DES\footnote{\url{https://www.darkenergysurvey.org/}} \citep{DES}, KiDS\footnote{\url{http://www.astro-wise.org/projects/KIDS/}} \citep{kilo-degree_survey}, Euclid\footnote{\url{http://sci.esa.int/euclid/}} \citep{Euclid}, and the Nancy Grace Roman Space Telescope \citep{NGRST} coming online soon, thousands of these lensed systems are expected to be found and an efficient image processing technique is required in order to process the large amount of scientific images that will be produced by either of these facilities. 

In order to study the detection efficiency of strongly lensed systems, the `Gravitational Lens Finding Challenge 1.0'\footnote{\url{http://metcalf1.difa.unibo.it/blf-portal/gg_challenge.html}} was launched in 2019 \citep{TheStrongGravitationalLensFindingChallenge}. The challenge consists of 100,000 objects, the aim being to detect whether each one is a gravitational lensed system or not. Many machine learning techniques are presented by \citet{TheStrongGravitationalLensFindingChallenge}, and this work aims to compare the techniques described in that paper with newer machine learning techniques, primarily \glspl{CNN}.

In the next section, we describe the framework developed and its features, followed by a description of the various methods implemented within it to tackle this problem. After this, we describe the dataset provided for the challenge, and what additional techniques were utilised to `expand' on this dataset. We then go on to describe what metrics are presented by our framework, and how methods are evaluated, and compare the performances achieved with those achieved in other works. We conclude the work by describing further improvements that can be implemented in order to make gravitational lensing detection methods more efficient and more accurate.

\subsection{LEXACTUM}

The framework developed in this work has been named \gls{LEXACTUM}.
The first of its main features are the Image Augmentation techniques described in Section~\ref{sec:image_augmentation} which can be toggled on or off to train for a greater number of epochs without overfitting.
Another feature is the modularity of the code, allowing for the rather easy development of new models, with very easy integration of new models into the pipeline.
Other features include the ability to set parameters from the command line. Examples of such parameters are the dataset path, whether to train a model or load one from disk, the name of the model (to train or load), the number of epochs to train for, the batch size, and whether to use image augmentation during training or not.
LEXACTUM also uses a custom `Data Generator', which loads and preprocesses images in batches with the CPU, while the GPU can train on the last batch of images. Apart from image augmentation during training, all images are normalised using ZScale \citepalias{zscale}. Like other components, the normalisation method can be easily swapped out for other techniques.
Furthermore, LEXACTUM comes with a `results' package, which scores the trained models and calculates several metrics, described in detail in Section~\ref{sec:results}.
Moreover, LEXACTUM saves trained models to disk, and also provides functionality for loading trained models.
Finally, the `visualise features' component was created, which allows for the viewing of the feature maps at every convolutional layer that a trained model is `looking at' during execution.

All of the architectures described in Section~\ref{sec:convolutional_neural_networks} were implemented in \gls{LEXACTUM}. All of these models were then trained from scratch, on both the Space and Ground datasets, using ZScale normalisation and image augmentation, for a varying number of epochs. As a starting point, all models were trained for 5 or 10, 25 or 50, 100, and 250 epochs. After that, if, say, a particular model obtained promising results, and didn't seem to be overfitting (judging by the loss and accuracy of the validation set) after 250 epochs, it would then be further trained for 500, or even 1,000 epochs.

\subsection{Literature Review}
\label{sec:literature_review}

\subsubsection{Conventional Methods}

The methods described in this subsection are not implemented in \gls{LEXACTUM}, and are only presented to give a broad view of what other methods exist for tackling this problem.

\paragraph{Visual Inspection}

\citet{visual_inspection_svm} go about this problem by visually inspecting and labelling each of the 100,000 images in each of the 2 datasets. Using their tool, \textsc{bigeye}, \citet{visual_inspection_svm} claim that they can label around 2,500 or 5,000 images an hour. The final score achieved by this solution was 0.804 for the Space set and 0.889 for the Ground set \citep{TheStrongGravitationalLensFindingChallenge}.
The score metric used is discussed in Section~\ref{sec:results}.

\paragraph{Arc-finders}

Arc-finders, such as \textsc{arcfinder} \citep{arcfinder} and \textsc{YattaLensLite} \citep{yatta_lens_lite}, attempt to detect lensing by looking for elongated structures, which are indicative of lensing. \textsc{arcfinder} achieves a score of 0.66 on the Space Set, whereas \textsc{YattaLensLite} achieves a score of 0.76 on the Space set and 0.82 on the Ground set \citep{TheStrongGravitationalLensFindingChallenge}.

\paragraph{Machine Learning (Pre-Selected Features)}

Such methods normally involve the creation of a feature space of features deemed to be relevant by an expert. The classification is then determined by a boundary, specified either by intuition or trial-and-error. \citet{visual_inspection_svm} attempted to solve this challenge with \textsc{Manchester-SVM}, an \gls{SVM} \citep{svm} based solution which achieved a score of 0.81 on the Space set and 0.93 on the Ground set. \citet{all}, on the other hand, use a \gls{HOG} \citep{hog} based approach in their solution, \textsc{all}, which scored 0.73 on the Space set and 0.84 on the Ground set \citep{TheStrongGravitationalLensFindingChallenge}.

\subsubsection{Convolutional Neural Networks}
\label{sec:convolutional_neural_networks}

\glspl{CNN} have shown to achieve very good results for both detection and recognition tasks in images and videos, among other applications.
A \gls{CNN} is a Neural Network that contains a Convolutional Layer. A convolutional layer `slides' a kernel (also referred to as a filter) over the input image, or the output from the previous convolutional layer, and computes the output as the convolution of the pixels the `sliding window' is currently over and the kernel.
Each Convolutional Layer has a number of filters, each of which can be described as a pattern detector.
The earlier layers extract geometric features, such as edges and corners, whereas deeper layers start to extract more sophisticated features, and are more capable of detecting objects such as eyes or noses \citep{lecun_cnn_architecture}.

The need for Convolutional Layers in \glspl{CNN} arises from the limitations of traditional Fully Connected Layers when dealing with images. One such limitation is that, for a 101x101 pixel image, one layer would have more than 10 thousand neurons, meaning one fully connected layer will thus have more than 100 million weights to be learnt. To put this value into perspective, from the \glspl{CNN} implemented in this work, the total number of weights ranges from around 100 thousand to around 6 million, for the entirety of each network.
One further limitation of Fully Connected Layers when dealing with 2-D images, or 3-D images when using images with more than one channel, is that when the images are flattened, most of the spatial correlation between pixels is lost. These ``local correlations'' are very important for the recognition of low-level features, such as edges \citep{lecun_convolutional_layers}.

All the techniques mentioned in this subsection are \gls{CNN} based, and have been implemented in \gls{LEXACTUM}. They have been implemented in Python\footnote{\url{https://www.python.org/}} using Keras\footnote{\url{https://keras.io/}}, a high-level \gls{API} for TensorFlow\footnote{\url{https://www.tensorflow.org/}}. All the source code and trained models mentioned in the results section are available on the GitHub repository \url{https://github.com/DanielMagro97/LEXACTUM}.

\paragraph{CAS Swinburne}
\label{sec:cas_swinburne}

This model was based on AlexNet \citep[][]{alex_net}. The input image is first passed through three convolutional layers, each followed by a \gls{ReLU} activation function and a max pooling layer. The output from the last max pool was put into two successive fully-connected layers, each followed by a dropout layer \citep{cas_swinburne, TheStrongGravitationalLensFindingChallenge}. This model is discussed in further detail in Appendix~\ref{sec:cas_swinburne_appendix}.

\paragraph{LASTRO EPFL}
\label{sec:lastro_epfl}

This model follows a somewhat similar architecture to that described in Section~\ref{sec:cas_swinburne}, however has significantly more layers.
This model starts off with 3 blocks, each block consisting of two consecutive convolutional layers, followed by a max pooling layer and a batch normalisation layer. The third block is followed by a dropout layer to reduce the possibility of overfitting.
Another pair of convolutional layers are added, each followed by a dropout layer.
The last layer's output is passed to a triple of fully-connected layers, which finally connect to a fully connected layer with a single neuron and a sigmoid activation \citep{lastro_epfl, TheStrongGravitationalLensFindingChallenge}. This model is discussed in further detail in Appendix~\ref{sec:lastro_epfl_appendix}.

\paragraph{CMU DeepLens}
\label{sec:cmu_deeplens}

Like the previously described models, this model is CNN based, however it is made up of `ResNet blocks'. A ResNet is a network where the input is passed through a series of convolutional layers, and the output is the addition of the original input and the output of the last convolution layer. This ``shortcut connection'' from the input of the block to the end of it tackles the `vanishing gradient problem', as it provides a `faster' route for the gradients from back propagation to reach the earlier layers.

In the CMU DeepLens model, two different types of `ResNet blocks' are used, one which keeps the original resolution of the image, and another which downsamples the image. In each case, the input of the `ResNet block' goes through three convolutional layers, and is summed with the original input to the block to create the aforementioned ``shortcut connection''.

The CMU DeepLens model starts by passing the input image to a convolutional layer, followed by 5 groups of 3 successive ResNet blocks. The output from the last block is passed through an Average Pooling layer, and the model's prediction is computed by a fully connected layer with one neuron and a sigmoid activation \citep{cmu_deeplens, TheStrongGravitationalLensFindingChallenge}. This model is discussed in further detail in Appendix~\ref{sec:cmu_deeplens_appendix}.

\paragraph{WSI-Net}
\label{sec:wsi_net}

The WSI-Net model described in this paper was originally used to first detect tumours in breast scans, and then classify them. The same model was used up to the point of detection, to detect the presence of a lens in an image. The original paper doesn't specify hyperparameter values, those presented in this paper are those found to produce the best results, empirically.

The model starts with a Convolutional Layer, followed by two ResNet blocks. These ResNet blocks used are the same as those described in Section~\ref{sec:cmu_deeplens}. This is followed by two blocks, each block consisting of a Convolutional Layer, a Batch Normalisation Layer, and a \gls{ReLU} activation. A Max Pooling Layer is added on next, followed by 2 Fully Connected Layers, the latter with 1 neuron and a sigmoid activation which determines the final classification \citep{wsi_net}. This model is discussed in further detail in Appendix~\ref{sec:wsi_net_appendix}.

\paragraph{LensFlow}
\label{sec:lens_flow}

In this model, the first operation carried out on the input image is an Average Pool. This is followed by a triple of `Convolutional Layer + Max Pool' pairs. The last max pool layer is fed into a Fully Connected layer. During training, this layer is followed by a dropout layer to reduce overfitting. The final output is obtained from a Fully Connected layer with 1 neuron, and a sigmoid activation \citep{lens_flow}. This architecture is discussed in further detail in Appendix~\ref{sec:lens_flow_appendix}.

\paragraph{Lens Finder}
\label{sec:lens_finder}

The LensFinder model has a relatively simplistic architecture, when compared to some of the solutions presented in this paper, however holds its weight with the score it obtains.
The paper doesn't state specific values for the hyper parameters of each layer in the model, the values presented here are what were found to work best, empirically.
The model starts with 2 blocks of `convolutional layer, max pooling layer, and \gls{ReLU} activation'. This is connected to a Fully Connected layer, which in turn connects to the final Fully Connected layer.
In the original paper, a softmax activation is used, however since this is a binary classification problem, only 1 neuron is used in this layer, and a sigmoid activation is used instead \citep{lens_finder}. This model is discussed in further detail in Appendix~\ref{sec:lens_finder_appendix}.

\section{Methodology}
\label{sec:methodology}

\subsection{The Datasets}

\begin{figure}
	\includegraphics[width=\columnwidth]{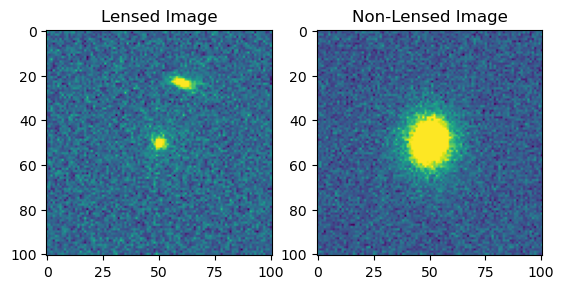}
    \caption{The image on the left is a random lensed image from the Space set, whereas the image on the right does not contain lensing. Reproduced from \citet{TheStrongGravitationalLensFindingChallenge}.}
    \label{fig:sample_space_set_images}
\end{figure}

\begin{figure}
	\includegraphics[width=\columnwidth]{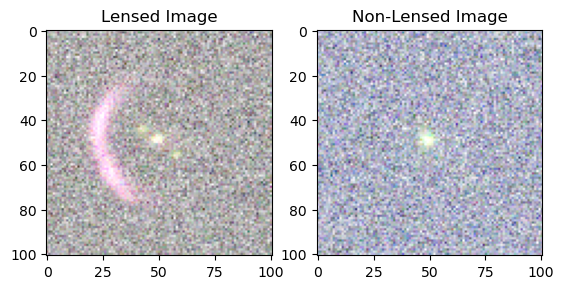}
    \caption{The image on the left is a random lensed image from the Ground set, whereas the image on the right does not contain lensing. Reproduced from \citet{TheStrongGravitationalLensFindingChallenge}.}
    \label{fig:sample_ground_set_images}
\end{figure}

Two separate labelled datasets of optical images were provided for this challenge, each with 100,000 simulated images. The first, called the `Space' dataset, is made up of single band (single channel) images which mimic data ``from a satellite survey such as \textit{Euclid}'' \citep{TheStrongGravitationalLensFindingChallenge}. The `Ground' dataset, on the other hand, was simulated such that it mimics a ground based survey, such as \gls{KiDS} \citep[][]{kilo-degree_survey}, where each image has 4 channels of data in the ``bands (I, G, R, U)'' \citep[][]{TheStrongGravitationalLensFindingChallenge}: Infrared (806~nm), Green (464~nm), Red (658~nm), Ultraviolet (365~nm) \citep{igru_bands}.
For each dataset, 20,000 of the 100,000 images were provided for training, whereas the other 80,000 were intended for evaluating and scoring the model.
The datasets can be downloaded from \url{http://metcalf1.difa.unibo.it/blf-portal/gg_challenge.html}, for this work, `Space set 1' and `Ground set 1' were used.
Fig.~\ref{fig:sample_space_set_images} shows an example of an image containing gravitational lensing, and another which doesn't, from the Space set. Similarly, Fig.~\ref{fig:sample_ground_set_images} shows the two cases from the Ground set.


\subsection{Image Augmentation}
\label{sec:image_augmentation}

20,000 training examples are provided for this challenge. In order to add diversity to the training set, and allow the model to train for a greater number of epochs without overfitting, Image Augmentation is employed. Image augmentation defines a set of transformations that can be applied to an image before it is passed to the neural network for training. It is important to note that this technique is only utilised for the training set, and not for validation or evaluation.

The image augmentation component utilises the `imgaug' library\footnote{\url{https://pypi.org/project/imgaug/}} to define 9 different transformations, of which a random amount are picked to be applied to the image. The transformations defined are the following:
\begin{itemize}
  \item A Vertical or Horizontal flipping of the image.
  \item A 90\textdegree, 180\textdegree or 270\textdegree Rotation of the image.
  \item A Translation of [-10\%, 10\%] of the image along the X and/or Y axes
  \item A Scaling of [0.75, 1] of the image along the X and/or Y axes
  \item A Shearing of [-20\%, 20\%] of the image along the X and/or Y axes
\end{itemize}

CAS Swinburne, LASTRO EPFL, CMU DeepLens, and WSI-Net \citep{cas_swinburne, lastro_epfl, cmu_deeplens, wsi_net} all utilise image augmentation during training, however the techniques used are generally limited to flips and rotations. One feature of this framework is that it offers those transformations, along with the other three mentioned previously, as a standard to any architecture added to it.

\section{Results}
\label{sec:results}

The metrics used in the paper by \citet{TheStrongGravitationalLensFindingChallenge} were the \gls{AUC}, {\tprzero} and {\tprten}.

The \gls{TPR} is the rate of instances correctly labelled as positive. The \gls{FPR} is the rate of instances incorrectly labelled as positive, and thus are actually negative.
The \gls{ROC} is a plot of the \gls{TPR} against the \gls{FPR} at various thresholds. Such a plot illustrates the performance of the model, where a curve which is close to the TPR=FPR diagonal would represent a model which is as effective as a coin flip, and a curve which very steeply approaches the value of TPR=1 represents a model that can achieve a high \gls{TPR} without labelling many False Positives.
The Area Under the \gls{ROC} (AUROC), or more simply the \gls{AUC}, is a convenient method of quantitatively comparing \glspl{ROC}.

The {\tprzero} is the highest \gls{TPR} achievable by the model, while keeping the \gls{FPR} equal to 0. Given the difficulty in achieving a {\tprzero} which is not 0, the {\tprten} is defined, which is similarly the highest \gls{TPR} achievable, while not classifying more than 10 false positives \citep[][]{TheStrongGravitationalLensFindingChallenge}.

The final metric that was recorded for this paper was the average execution time of each model. This was obtained by recording the length of time it took for the already trained model to evaluate the test set. This was then divided by the number of images in the test set to obtain the average execution time for one image. The execution times for the same model trained for different numbers of epochs were averaged out, as they are still the same model. Furthermore, any times where the time was significantly different than the rest (outliers) were ignored, and not included in the average.

\subsection{Space Set Results}

Results obtained on the Space dataset are shown in Table~\ref{tab:space_results}.
The best \gls{TPR} achieved was 0.8738 by CMU DeepLens when trained for just 25 epochs.
The best \gls{FPR} achieved was 0.0042 by Lastro EPFL when trained for 5 epochs.
The best \gls{AUC} was 0.9343, by CMU DeepLens when trained for 500 epochs. In \citet{TheStrongGravitationalLensFindingChallenge}'s paper, the best AUC for the Space set was 0.93 by LASTRO EPFL, whereas the implementation of CMU DeepLens scored 0.92.
The best {\tprzero} was 0.2411, by CAS Swinburne when trained for 50 epochs. In \citet{TheStrongGravitationalLensFindingChallenge}'s paper, the best {\tprzero} for the Space set was 0.22, by CMU DeepLens.
The best {\tprten} was 0.4211, by WSI Net when trained for 250 epochs. In \citet{TheStrongGravitationalLensFindingChallenge}'s paper, the best {\tprten} for the Space set was 0.36, by GAMOCLASS, another \gls{CNN} based solution. This is a very interesting finding, as a ResNet based network which was not included in the \citet{TheStrongGravitationalLensFindingChallenge} paper, achieved a significantly higher score than that in the paper.

\begin{table*}
\caption{Table showing the \gls{TPR}, \gls{FPR}, \gls{AUC}, {\tprzero}, {\tprten}, and average execution time for 6 different models, as described in Section~\ref{sec:convolutional_neural_networks}, trained for a various number of epochs on the Space dataset.
Columns marked with an * indicate the score achieved by the model in \citet{TheStrongGravitationalLensFindingChallenge}. Values in these columns marked in green indicate better performance compared to our implementations in \gls{LEXACTUM}, whereas values in red indicate worse performance.}
\label{tab:space_results}
\begin{tabular}{l|llllll|lll|l}
\hline
Model Name & \begin{tabular}[c]{@{}l@{}}No. of\\ Training Epochs\end{tabular} & TPR & FPR & AUC & \tprzero & \tprten & AUC* & {\tprzero}* & {\tprten}* & \begin{tabular}[c]{@{}l@{}}Avg. Execution Time\\ per Image (seconds)\end{tabular} \\ \hline
CAS Swinburne & 5 & 0.5250 & 0.0603 & 0.8489 & 0.1531 & 0.1861 & \multicolumn{3}{l|}{} &  \\
 & 10 & 0.5517 & 0.1077 & 0.8171 & 0.1054 & 0.1509 & \multicolumn{3}{l|}{} &  \\
 & 25 & 0.7221 & 0.1178 & 0.8870 & 0.0000 & 0.2705 & \multicolumn{3}{l|}{} &  \\
 & 50 & 0.6252 & 0.0461 & 0.8894 & \cellcolor[HTML]{C6EFCE}{\color[HTML]{006100} 0.2411} & 0.3000 & \multicolumn{3}{l|}{} &  \\
 & 75 & 0.6503 & 0.0474 & 0.8963 & 0.0000 & 0.3221 & \multicolumn{3}{l|}{} &  \\
 & 100 & 0.6604 & 0.0591 & 0.8915 & 0.0000 & 0.3016 & \multicolumn{3}{l|}{} &  \\
 & 500 & 0.6551 & 0.0295 & 0.9086 & 0.0000 & 0.3602 & \multicolumn{3}{l|}{\multirow{-7}{*}{N/A}} & \multirow{-7}{*}{0.0124} \\ \hline
Lastro EPFL & 5 & 0.3507 & 0.0042 & 0.8641 & 0.1539 & 0.2112 &  &  &  &  \\
 & 10 & 0.7302 & 0.3543 & 0.7825 & 0.1894 & 0.2455 &  &  &  &  \\
 & 50 & 0.6650 & 0.0287 & 0.9132 & 0.2107 & 0.3823 &  &  &  &  \\
 & 250 & 0.7937 & 0.0687 & 0.9322 & 0.0000 & 0.2268 & \multirow{-4}{*}{0.93} & \multirow{-4}{*}{\textcolor{red}{0.00}} & \multirow{-4}{*}{\textcolor{red}{0.08}} & \multirow{-4}{*}{0.0061} \\ \hline
CMU Deeplens & 5 & 0.6056 & 0.1539 & 0.7984 & 0.0000 & 0.1206 &  &  &  &  \\
 & 10 & 0.8268 & 0.2880 & 0.8710 & 0.0000 & 0.2309 &  &  &  &  \\
 & 25 & 0.8738 & 0.2726 & 0.9113 & 0.0000 & 0.0000 &  &  &  &  \\
 & 50 & 0.7570 & 0.0628 & 0.9243 & 0.0000 & 0.4073 &  &  &  &  \\
 & 100 & 0.8170 & 0.1321 & 0.9226 & 0.0000 & 0.0000 &  &  &  &  \\
 & 250 & 0.7592 & 0.0436 & 0.9291 & 0.0000 & 0.0000 &  &  &  &  \\
 & 500 & 0.7952 & 0.0626 & \cellcolor[HTML]{C6EFCE}{\color[HTML]{006100} 0.9343} & 0.0000 & 0.0000 &  &  &  &  \\
 & 1000 & 0.8611 & 0.1634 & 0.9303 & 0.0000 & 0.0000 & \multirow{-8}{*}{\textcolor{red}{0.92}} & \multirow{-8}{*}{\textcolor{green}{0.22}} & \multirow{-8}{*}{\textcolor{red}{0.29}} & \multirow{-8}{*}{0.0061} \\ \hline
WSI Net & 5 & 0.7132 & 0.2955 & 0.7935 & 0.0000 & 0.0000 & \multicolumn{3}{l|}{} &  \\
 & 10 & 0.5437 & 0.0187 & 0.8867 & 0.1799 & 0.2934 & \multicolumn{3}{l|}{} &  \\
 & 50 & 0.7888 & 0.1194 & 0.9115 & 0.0000 & 0.0000 & \multicolumn{3}{l|}{} &  \\
 & 100 & 0.7348 & 0.0624 & 0.9069 & 0.0000 & 0.3976 & \multicolumn{3}{l|}{} &  \\
 & 250 & 0.7255 & 0.0531 & 0.9083 & 0.0000 & \cellcolor[HTML]{C6EFCE}{\color[HTML]{006100} 0.4211} & \multicolumn{3}{l|}{\multirow{-5}{*}{N/A}} & \multirow{-5}{*}{0.0055} \\ \hline
Lens Flow & 5 & 0.6508 & 0.1520 & 0.8389 & 0.0728 & 0.1260 & \multicolumn{3}{l|}{} &  \\
 & 25 & 0.6431 & 0.0726 & 0.8799 & 0.1903 & 0.2704 & \multicolumn{3}{l|}{} &  \\
 & 100 & 0.6780 & 0.0636 & 0.8963 & 0.0000 & 0.3379 & \multicolumn{3}{l|}{} &  \\
 & 250 & 0.7384 & 0.0889 & 0.9046 & 0.0000 & 0.3632 & \multicolumn{3}{l|}{\multirow{-4}{*}{N/A}} & \multirow{-4}{*}{0.0054} \\ \hline
Lens Finder & 5 & 0.4915 & 0.1001 & 0.8038 & 0.0885 & 0.1056 & \multicolumn{3}{l|}{} &  \\
 & 25 & 0.6203 & 0.0663 & 0.8739 & 0.2103 & 0.2395 & \multicolumn{3}{l|}{} &  \\
 & 100 & 0.6912 & 0.0855 & 0.8857 & 0.0000 & 0.2721 & \multicolumn{3}{l|}{} &  \\
 & 250 & 0.7651 & 0.1062 & 0.9056 & 0.0000 & 0.3739 & \multicolumn{3}{l|}{\multirow{-4}{*}{N/A}} & \multirow{-4}{*}{0.0197} \\ \hline
\end{tabular}
\end{table*}

\subsection{Ground Set Results}

Results obtained on the Ground dataset are shown in Table~\ref{tab:ground_results}.
The best \gls{TPR} achieved was 0.9333 by CMU DeepLens when trained for 100 epochs.
The best \gls{FPR} achieved was 0.0232, by CMU DeepLens when trained for 25 epochs.
The best \gls{AUC} was 0.9870, by CMU DeepLens when trained for 150 epochs. In \citet{TheStrongGravitationalLensFindingChallenge}'s paper, the best AUC for the Ground set was 0.98, also by CMU DeepLens.
The best {\tprzero} was 0.6046, by CMU DeepLens when trained for 50 epochs. In \citet{TheStrongGravitationalLensFindingChallenge}'s paper, the best {\tprzero} for the Ground set was 0.22, by Manchester SVM.
The best {\tprten} was 0.7042, again by CMU DeepLens when trained for 150 epochs. In \citet{TheStrongGravitationalLensFindingChallenge}'s paper, the best {\tprten} for the Ground set was 0.45, also by CMU DeepLens. This is another very significant improvement, using essentially the same network as specified in the original paper. The only differences are the usage of slightly different image augmentation techniques, which possibly allowed our model to train for more epochs without overfitting. As we trained for up to 250 epochs, we were able to find more optimal weights at 150 epochs, whereas in the original work, the model was trained up to 120 epochs \citep[][]{cmu_deeplens}.


All the models described were trained and evaluated from scratch again, for both the Space and Ground dataset, using a different split of the training, validation and test sets (same ratio, different selection). This was done to evaluate the consistency of the results. It resulted that, for the Space set, between the two runs, the mean change between to runs of the same model with the same parameters was 0.96\%, with the greatest change between any two runs being 2.97\%. For the Ground set, the mean change was 0.39\%, with the greatest change being 2.99\%.

\begin{table*}
\caption{Table showing the \gls{TPR}, \gls{FPR}, \gls{AUC}, {\tprzero}, {\tprten}, and average execution time for 6 different models, as described in Section~\ref{sec:convolutional_neural_networks}, trained for a various number of epochs on the Ground dataset.
Columns marked with an * indicate the score achieved by the model in \citet{TheStrongGravitationalLensFindingChallenge}. Values in these columns marked in green indicate better performance compared to our implementations in \gls{LEXACTUM}, whereas values in red indicate worse performance.}
\label{tab:ground_results}
\begin{tabular}{l|llllll|lll|l}
\hline
Model Name & \begin{tabular}[c]{@{}l@{}}No. of\\ Training Epochs\end{tabular} & TPR & FPR & AUC & \tprzero & \tprten & AUC* & {\tprzero}* & {\tprten}* & \begin{tabular}[c]{@{}l@{}}Avg. Execution Time\\ per Image (seconds)\end{tabular} \\ \hline
CAS   Swinburne & 10 & 0.8779 & 0.1077 & 0.9608 & 0.0000 & 0.0000 &  &  &  &  \\
 & 50 & 0.8995 & 0.0944 & 0.9720 & 0.0000 & 0.0000 &  &  &  &  \\
 & 100 & 0.8565 & 0.0406 & 0.9742 & 0.0000 & 0.0000 &  &  &  &  \\
 & 250 & 0.8726 & 0.0429 & 0.9758 & 0.0000 & 0.0000 & \multirow{-4}{*}{\textcolor{red}{0.96}} & \multirow{-4}{*}{\textcolor{green}{0.02}} & \multirow{-4}{*}{\textcolor{green}{0.08}} & \multirow{-4}{*}{0.0469} \\ \hline
Lastro EPFL & 50 & 0.9073 & 0.0536 & 0.9824 & 0.0000 & 0.5133 &  &  &  &  \\
 & 100 & 0.9110 & 0.0482 & 0.9844 & 0.0000 & 0.5504 &  &  &  &  \\
 & 250 & 0.9197 & 0.0489 & 0.9862 & 0.0000 & 0.0000 & \multirow{-3}{*}{\textcolor{red}{0.97}} & \multirow{-3}{*}{\textcolor{green}{0.07}} & \multirow{-3}{*}{\textcolor{red}{0.11}} & \multirow{-3}{*}{0.0429} \\ \hline
CMU Deeplens & 25 & 0.7733 & 0.0232 & 0.9588 & 0.0000 & 0.3840 &  &  &  &  \\
 & 50 & 0.9138 & 0.0568 & 0.9825 & \cellcolor[HTML]{C6EFCE}{\color[HTML]{006100} 0.6046} & 0.6827 &  &  &  &  \\
 & 75 & 0.9026 & 0.0550 & 0.9804 & 0.0000 & 0.6536 &  &  &  &  \\
 & 100 & 0.9333 & 0.0660 & 0.9851 & 0.0000 & 0.6673 &  &  &  &  \\
 & 150 & 0.9205 & 0.0445 & \cellcolor[HTML]{C6EFCE}{\color[HTML]{006100} 0.9870} & 0.0000 & \cellcolor[HTML]{C6EFCE}{\color[HTML]{006100} 0.7042} &  &  &  &  \\
 & 250 & 0.8593 & 0.0858 & 0.9570 & 0.0000 & 0.0000 & \multirow{-6}{*}{\textcolor{red}{0.98}} & \multirow{-6}{*}{\textcolor{red}{0.09}} & \multirow{-6}{*}{\textcolor{red}{0.45}} & \multirow{-6}{*}{0.0594} \\ \hline
WSI Net & 50 & 0.8560 & 0.0589 & 0.9620 & 0.0000 & 0.0000 & \multicolumn{3}{l|}{} &  \\
 & 100 & 0.8218 & 0.0301 & 0.9710 & 0.0000 & 0.5347 & \multicolumn{3}{l|}{} &  \\
 & 250 & 0.9127 & 0.0864 & 0.9742 & 0.0000 & 0.0000 & \multicolumn{3}{l|}{\multirow{-3}{*}{N/A}} & \multirow{-3}{*}{0.0231} \\ \hline
Lens Flow & 50 & 0.8784 & 0.0744 & 0.9708 & 0.0000 & 0.5101 & \multicolumn{3}{l|}{} &  \\
 & 100 & 0.8831 & 0.0738 & 0.9726 & 0.0000 & 0.5648 & \multicolumn{3}{l|}{} &  \\
 & 250 & 0.9006 & 0.0733 & 0.9758 & 0.0000 & 0.0000 & \multicolumn{3}{l|}{\multirow{-3}{*}{N/A}} & \multirow{-3}{*}{0.0349} \\ \hline
Lens Finder & 50 & 0.8556 & 0.0648 & 0.9665 & 0.0000 & 0.4442 & \multicolumn{3}{l|}{} &  \\
 & 100 & 0.8938 & 0.0805 & 0.9718 & 0.0000 & 0.5664 & \multicolumn{3}{l|}{} &  \\
 & 250 & 0.8997 & 0.0880 & 0.9671 & 0.0000 & 0.0000 & \multicolumn{3}{l|}{\multirow{-3}{*}{N/A}} & \multirow{-3}{*}{0.0293} \\ \hline
\end{tabular}
\end{table*}

\subsection{The Importance of Image Augmentation}

From Table~\ref{tab:space_results}, the results for CMU DeepLens when trained for 250 epochs with the Space dataset using image augmentation are an \gls{AUC} of 0.9291, \gls{TPR} of 0.7592 and a \gls{FPR} of 0.0436.
To demonstrate the effectiveness of image augmentation, the same model was trained with the same dataset, and parameters, only without image augmentation. The \gls{AUC} obtrained was 0.8800, the \gls{TPR} 0.7103 and the \gls{FPR} 0.1003.
The accuracy of the model (without image augmentation) on the training data during training can be seen rising epoch after epoch, and reaches 0.9996. On the other hand, the accuracy of the model on the validation set after 250 epochs was only 0.8156. The model obtains such a score as early as the 15\textsuperscript{th} epoch, showing that the accuracy fails to improve and, thus, that the model is overfitting.
When using image augmentation during training, after the same number of epochs, the model `only' reaches an accuracy of 0.8989 on the training set, however manages a, relatively, impressive 0.8813 accuracy on the validation set. By the 15\textsuperscript{th} epoch, this model has already achieved a validation accuracy of 0.8333, however manages to further improve on this, and as mentioned climbs to 0.88813.

\subsection{Visualising and Interpreting Features Extracted by Convolutional Layers}

\begin{figure}
	\includegraphics[width=\columnwidth]{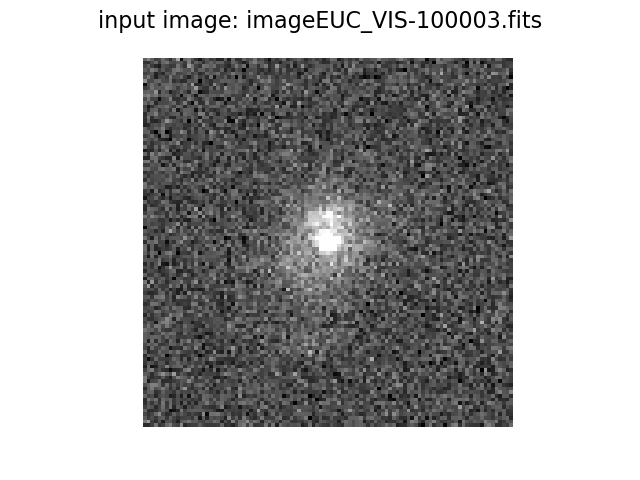}
    \caption{This is the input image, `imageEUC\_VIS-100003.fits', used to visualise the features extracted by the CMU DeepLens model which was trained for 500 epochs.}
    \label{fig:input_image}
\end{figure}

The `visualise features' component makes it possible to visualise the outputs of each convolutional layer, for any chosen model given any image. Since it scored the highest, the `space\_cmu\_deeplens\_500epochs.h5' model was executed with a random image from the dataset as an input, shown in Fig.~\ref{fig:input_image}. The features extracted by each convolutional layer were visualised and will be interpreted in this section.

\begin{figure}
	\includegraphics[width=\columnwidth]{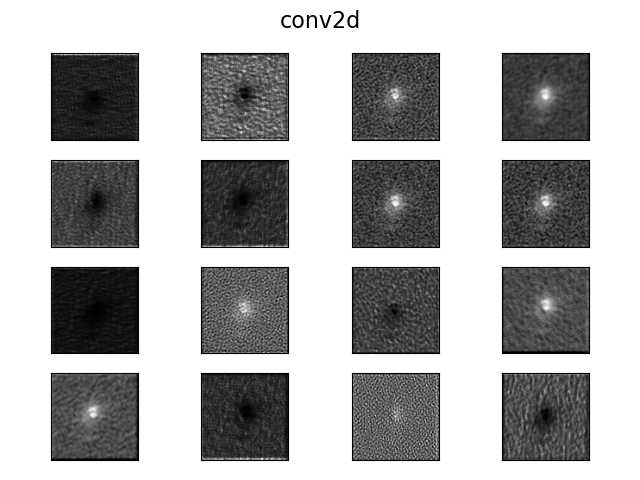}
    \caption{This is a visualisation of the features extracted by the first convolutional layer of a CMU DeepLens model which was trained for 500 epochs.}
    \label{fig:conv2d}
\end{figure}

A sample of the features extracted by the first convolutional layer are shown in Fig.~\ref{fig:conv2d}. At this stage the original image is still very clear in the extracted features, which is to be expected as at this stage the model is still in the process of extracting fine details from the image. For instance, the different features show the model's efforts to emphasise certain details (that it has learnt are important and relevant) by changing the brightness, the separation from the foreground object to the background, and the emphasis on the boundary between them, to name a few.

\begin{figure}
	\includegraphics[width=\columnwidth]{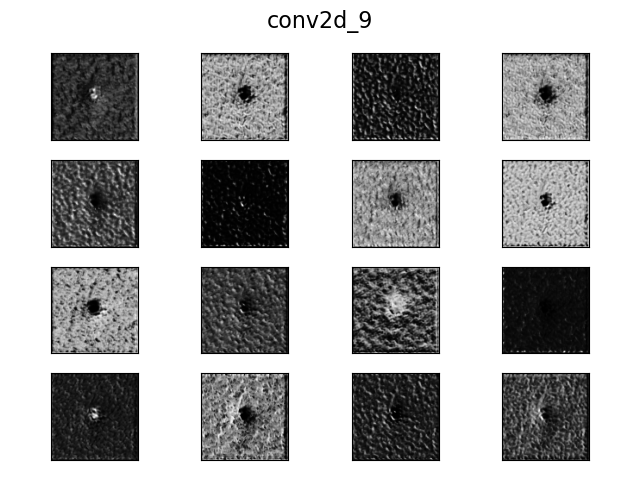}
    \caption{This is a visualisation of the features extracted by the last convolutional layer of the first `3 ResNet block' of a CMU DeepLens model which was trained for 500 epochs.}
    \label{fig:conv2d_9}
\end{figure}

Another sample of features extracted by the last convolutional layer of the first `3 ResNet block' are shown in Fig.~\ref{fig:conv2d_9}. Similarly to the first convolutional stage, the original image is still quite visible in the features extracted by this layer. At this stage the model is still looking at fine details in the image.

In Fig.~\ref{fig:conv2d_19,29,39,49}, a sample of features extracted from the last convolutional layer of the remaining `3 ResNet blocks' are shown. With each successive convolutional layer, the features extracted show less and less detail, with the features becoming increasingly difficult to interpret. \citet{visualise_filters} explains that this is due to the model extracting more abstract features in the deeper layers which show ``more general concepts'' that make it easier for the model to make a classification.
\begin{figure}
	\includegraphics[width=\columnwidth]{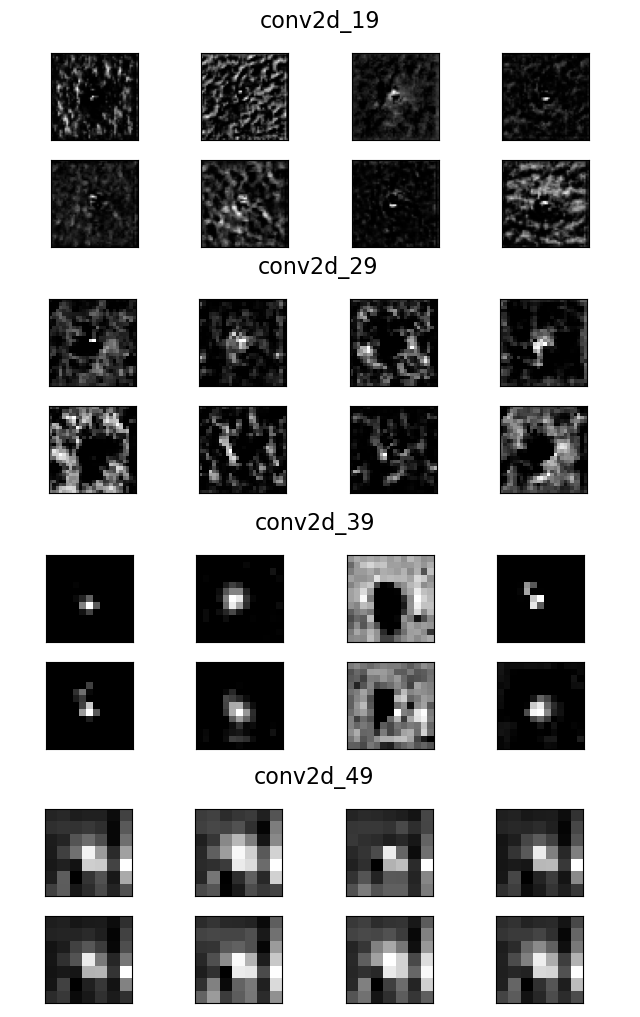}
    \caption{This is a visualisation of the features extracted by the last convolutional layer of the remaining `3 ResNet blocks' of a CMU DeepLens model which was trained for 500 epochs.}
    \label{fig:conv2d_19,29,39,49}
\end{figure}

\section{Conclusions}

It is fair to say that the developed framework, \gls{LEXACTUM}, makes it significantly easier to develop new network architectures, or apply existing ones, to the Gravitational Lensing problem, with its readily available image normalisation and image augmentation features. Furthermore, \gls{LEXACTUM} provides standard metrics to evaluate the performance of the models, along with ready made functionality for saving, and loading, trained models.

In this paper, some solutions which were already tried in the original paper by \citet{TheStrongGravitationalLensFindingChallenge} were reimplemented with image augmentation, and in some cases achieved significantly better results than what was reported. Furthermore, new techniques were implemented and used, in particular WSI-Net, which registered 17\% improvement in {\tprten} over the winning solution in the original paper for the Space dataset. A 56\% improvement in {\tprten} was also registered over the winning solution for the Ground set by CMU DeepLens. CMU DeepLens also registered a very impressive 175\% improvement over the {\tprzero} for the Ground set.

The work done here applies data preprocessing, in particular augmentation techniques, for extended training of models all the while avoiding overfitting the model to the training data. Furthermore, new techniques which were previously applied to other fields were applied to this problem, with the results obtained confirming the adaptability of \glspl{CNN}. Ultimately, this work further proves the effectiveness of \glspl{CNN} based techniques for astronomical data problems.

\subsection{Future Work}

It would be interesting to experiment with applying an Elliptical Hough Transform to the images as a preprocessing step, as this may make it easier for the models to locate the features which determine whether an image is classified as being lenses or not. \citet{elliptical_hough_transform} attempt to do something similar, however for their use case, they noted that it was only able to detect larger features. With this in mind, perhaps the output of the transform could be fed to the networks as an additional channel, rather than replacing the original image.

One other task that could be carried out to possibly maximise the performance of the existing models, is to run hyper-parameter optimisation. A module could possibly be added to \gls{LEXACTUM} which does this automatically with minimal configuration.

Further image augmentation techniques could also be tested, which would possibly allow the networks to train for an even greater number of epochs without overfitting.

Lastly, new network architectures can also be assessed. ResNet based networks showed very promising results for this particular problem.

\section*{Acknowledgements}

I would like to express my sincere gratitude to the \gls{OACT} and \gls{INAF} for their financial support.

\section*{Data availability}

The datasets used for training and evaluating the \glspl{CNN} are publicly available on the `Gravitational Lens Finding Challenge 1.0' webpage, \url{http://metcalf1.difa.unibo.it/blf-portal/gg_challenge.html}.
All the code written for \gls{LEXACTUM} is also publicly available on the GitHub repository \url{https://github.com/DanielMagro97/LEXACTUM}, under the GNU General Public License v3.0\footnote{\url{https://www.gnu.org/licenses/gpl-3.0.html}}.
The weights files for the trained models have also been made available on Zenodo \url{https://doi.org/10.5281/zenodo.4299924} and Google Drive \url{https://drive.google.com/drive/folders/1qn03htSDz-0aB6jRWbKmDk4QBz0epSmS?usp=sharing} \citep{lexactum_trained_model_weights}.




\def\UrlBreaks{\do\/\do-}

\bibliographystyle{mnras}
\bibliography{references}

\begin{thebibliography}{}
\makeatletter
\relax
\def\mn@urlcharsother{\let\do\@makeother \do\$\do\&\do\#\do\^\do\_\do\%\do\~}
\def\mn@doi{\begingroup\mn@urlcharsother \@ifnextchar [ {\mn@doi@}
  {\mn@doi@[]}}
\def\mn@doi@[#1]#2{\def\@tempa{#1}\ifx\@tempa\@empty \href
  {http://dx.doi.org/#2} {doi:#2}\else \href {http://dx.doi.org/#2} {#1}\fi
  \endgroup}
\def\mn@eprint#1#2{\mn@eprint@#1:#2::\@nil}
\def\mn@eprint@arXiv#1{\href {http://arxiv.org/abs/#1} {{\tt arXiv:#1}}}
\def\mn@eprint@dblp#1{\href {http://dblp.uni-trier.de/rec/bibtex/#1.xml}
  {dblp:#1}}
\def\mn@eprint@#1:#2:#3:#4\@nil{\def\@tempa {#1}\def\@tempb {#2}\def\@tempc
  {#3}\ifx \@tempc \@empty \let \@tempc \@tempb \let \@tempb \@tempa \fi \ifx
  \@tempb \@empty \def\@tempb {arXiv}\fi \@ifundefined
  {mn@eprint@\@tempb}{\@tempb:\@tempc}{\expandafter \expandafter \csname
  mn@eprint@\@tempb\endcsname \expandafter{\@tempc}}}

\bibitem[\protect\citeauthoryear{Alard}{Alard}{2006}]{arcfinder}
Alard C.,  2006, arXiv preprint astro-ph/0606757

\bibitem[\protect\citeauthoryear{Avestruz, Li, Zhu, Lightman, Collett  \&
  Luo}{Avestruz et~al.}{2019}]{all}
Avestruz C.,  Li N.,  Zhu H.,  Lightman M.,  Collett T.~E.,   Luo W.,  2019,
  \mn@doi [The Astrophysical Journal] {10.3847/1538-4357/ab16d9}, 877, 58

\bibitem[\protect\citeauthoryear{Binney \& Merrifield}{Binney \&
  Merrifield}{1998}]{igru_bands}
Binney J.,  Merrifield M.,  1998, Galactic Astronomy.
Princeton University Press

\bibitem[\protect\citeauthoryear{Blake, Abdalla, Bridle  \& Rawlings}{Blake
  et~al.}{2004}]{SKA}
Blake C.,  Abdalla F.,  Bridle S.,   Rawlings S.,  2004, \mn@doi [New Astronomy
  Reviews] {https://doi.org/10.1016/j.newar.2004.09.045}, 48, 1063

\bibitem[\protect\citeauthoryear{Bolton, Burles, Koopmans, Treu  \&
  Moustakas}{Bolton et~al.}{2006}]{SLACS}
Bolton A.~S.,  Burles S.,  Koopmans L. V.~E.,  Treu T.,   Moustakas L.~A.,
  2006, \mn@doi [The Astrophysical Journal] {10.1086/498884}, 638, 703

\bibitem[\protect\citeauthoryear{Brownlee}{Brownlee}{2019}]{visualise_filters}
Brownlee J.,  2019, How to Visualize Filters and Feature Maps in Convolutional
  Neural Networks, \url
  {https://machinelearningmastery.com/how-to-visualize-filters-and-feature-maps-in-convolutional-neural-networks/}

\bibitem[\protect\citeauthoryear{Dalal \& Triggs}{Dalal \& Triggs}{2005}]{hog}
Dalal N.,  Triggs B.,  2005, in 2005 IEEE computer society conference on
  computer vision and pattern recognition (CVPR'05). pp 886--893

\bibitem[\protect\citeauthoryear{Dressler et~al.,}{Dressler
  et~al.}{2012}]{NGRST}
Dressler A.,  et~al., 2012, Exploring the NRO Opportunity for a Hubble-sized
  Wide-field Near-IR Space Telescope -- NEW WFIRST (\mn@eprint {arXiv}
  {1210.7809})

\bibitem[\protect\citeauthoryear{{Hartley}, {Flamary}, {Jackson}, {Tagore}  \&
  {Metcalf}}{{Hartley} et~al.}{2017}]{visual_inspection_svm}
{Hartley} P.,  {Flamary} R.,  {Jackson} N.,  {Tagore} A.~S.,   {Metcalf} R.~B.,
   2017, \mn@doi [\mnras] {10.1093/mnras/stx1733}, \href
  {https://ui.adsabs.harvard.edu/abs/2017MNRAS.471.3378H} {471, 3378}

\bibitem[\protect\citeauthoryear{{Jacobs}, {Glazebrook}, {Collett}, {More}  \&
  {McCarthy}}{{Jacobs} et~al.}{2017}]{cas_swinburne}
{Jacobs} C.,  {Glazebrook} K.,  {Collett} T.,  {More} A.,   {McCarthy} C.,
  2017, \mn@doi [\mnras] {10.1093/mnras/stx1492}, \href
  {https://ui.adsabs.harvard.edu/abs/2017MNRAS.471..167J} {471, 167}

\bibitem[\protect\citeauthoryear{Kochanek, Falco, Impey, Leh{\'a}r, McLeod  \&
  Rix}{Kochanek et~al.}{1999}]{castles_survey}
Kochanek C.,  Falco E.,  Impey C.,  Leh{\'a}r J.,  McLeod B.,   Rix H.-W.,
  1999, in AIP Conference Proceedings. pp 163--175

\bibitem[\protect\citeauthoryear{Krizhevsky, Sutskever  \& Hinton}{Krizhevsky
  et~al.}{2017}]{alex_net}
Krizhevsky A.,  Sutskever I.,   Hinton G.~E.,  2017, \mn@doi [Commun. ACM]
  {10.1145/3065386}, 60, 84–90

\bibitem[\protect\citeauthoryear{{LSST Science Collaboration} et~al.,}{{LSST
  Science Collaboration} et~al.}{2009}]{LSST}
{LSST Science Collaboration} et~al., 2009, LSST Science Book, Version 2.0
  (\mn@eprint {arXiv} {0912.0201})

\bibitem[\protect\citeauthoryear{{Lanusse}, {Ma}, {Li}, {Collett}, {Li},
  {Ravanbakhsh}, {Mandelbaum}  \& {P{\'o}czos}}{{Lanusse}
  et~al.}{2018}]{cmu_deeplens}
{Lanusse} F.,  {Ma} Q.,  {Li} N.,  {Collett} T.~E.,  {Li} C.-L.,  {Ravanbakhsh}
  S.,  {Mandelbaum} R.,   {P{\'o}czos} B.,  2018, \mn@doi [\mnras]
  {10.1093/mnras/stx1665}, \href
  {https://ui.adsabs.harvard.edu/abs/2018MNRAS.473.3895L} {473, 3895}

\bibitem[\protect\citeauthoryear{Laureijs et~al.,}{Laureijs
  et~al.}{2011}]{Euclid}
Laureijs R.,  et~al., 2011, Euclid Definition Study Report (\mn@eprint {arXiv}
  {1110.3193})

\bibitem[\protect\citeauthoryear{LeCun, Boser, Denker, Henderson, Howard,
  Hubbard  \& Jackel}{LeCun et~al.}{1989}]{lecun_cnn_architecture}
LeCun Y.,  Boser B.,  Denker J.,  Henderson D.,  Howard R.,  Hubbard W.,
  Jackel L.,  1989, Advances in neural information processing systems, 2, 396

\bibitem[\protect\citeauthoryear{LeCun, Bottou, Bengio  \& Haffner}{LeCun
  et~al.}{1998}]{lecun_convolutional_layers}
LeCun Y.,  Bottou L.,  Bengio Y.,   Haffner P.,  1998, Proceedings of the IEEE,
  86, 2278

\bibitem[\protect\citeauthoryear{Magro, Zarb~Adami, DeMarco, Riggi  \&
  Sciacca}{Magro et~al.}{2020}]{lexactum_trained_model_weights}
Magro D.,  Zarb~Adami K.,  DeMarco A.,  Riggi S.,   Sciacca E.,  2020, LEXACTUM
  Trained Model Weights, \mn@doi{10.5281/zenodo.4299924}, \url
  {https://doi.org/10.5281/zenodo.4299924}

\bibitem[\protect\citeauthoryear{{Metcalf, R. B.} et~al.,}{{Metcalf, R. B.}
  et~al.}{2019}]{TheStrongGravitationalLensFindingChallenge}
{Metcalf, R. B.} et~al., 2019, \mn@doi [A\&A] {10.1051/0004-6361/201832797},
  625, A119

\bibitem[\protect\citeauthoryear{More, Cabanac, More, Alard, Limousin, Kneib,
  Gavazzi  \& Motta}{More et~al.}{2012}]{SL2S}
More A.,  Cabanac R.,  More S.,  Alard C.,  Limousin M.,  Kneib J.-P.,  Gavazzi
  R.,   Motta V.,  2012, \mn@doi [The Astrophysical Journal]
  {10.1088/0004-637x/749/1/38}, 749, 38

\bibitem[\protect\citeauthoryear{Myers et~al.,}{Myers et~al.}{2003}]{CLASS}
Myers S.~T.,  et~al., 2003, \mn@doi [Monthly Notices of the Royal Astronomical
  Society] {10.1046/j.1365-8711.2003.06256.x}, 341, 1

\bibitem[\protect\citeauthoryear{{National Optical Astronomy
  Observatory}}{{National Optical Astronomy Observatory}}{1997}]{zscale}
{National Optical Astronomy Observatory} 1997, IRAF (Image Reduction and
  Analysis Facility) Display Help Page, \url
  {https://iraf.net/irafhelp.php?val=display}

\bibitem[\protect\citeauthoryear{Negrello et~al.,}{Negrello
  et~al.}{2016}]{H-ATLAS}
Negrello M.,  et~al., 2016, \mn@doi [Monthly Notices of the Royal Astronomical
  Society] {10.1093/mnras/stw2911}, 465, 3558

\bibitem[\protect\citeauthoryear{Ni, Liu, Wang, Wang, Zhou  \& Qian}{Ni
  et~al.}{2019}]{wsi_net}
Ni H.,  Liu H.,  Wang K.,  Wang X.,  Zhou X.,   Qian Y.,  2019, in Suk H.-I.,
  Liu M.,  Yan P.,   Lian C.,  eds, Machine Learning in Medical Imaging.
  Springer International Publishing, Cham, pp 36--44

\bibitem[\protect\citeauthoryear{Pearson, Pennock  \& Robinson}{Pearson
  et~al.}{2018}]{lens_finder}
Pearson J.,  Pennock C.,   Robinson T.,  2018, \mn@doi [Emergent Scientist]
  {10.1051/emsci/2017010}, 2, 1

\bibitem[\protect\citeauthoryear{Pourrahmani, Nayyeri  \& Cooray}{Pourrahmani
  et~al.}{2018}]{lens_flow}
Pourrahmani M.,  Nayyeri H.,   Cooray A.,  2018, \mn@doi [The Astrophysical
  Journal] {10.3847/1538-4357/aaae6a}, 856, 68

\bibitem[\protect\citeauthoryear{{Schaefer}, {Geiger}, {Kuntzer}  \&
  {Kneib}}{{Schaefer} et~al.}{2018}]{lastro_epfl}
{Schaefer} C.,  {Geiger} M.,  {Kuntzer} T.,   {Kneib} J.~P.,  2018, \mn@doi
  [\aap] {10.1051/0004-6361/201731201}, \href
  {https://ui.adsabs.harvard.edu/abs/2018A&A...611A...2S} {611, A2}

\bibitem[\protect\citeauthoryear{{Sonnenfeld} et~al.,}{{Sonnenfeld}
  et~al.}{2018}]{yatta_lens_lite}
{Sonnenfeld} A.,  et~al., 2018, \mn@doi [\pasj] {10.1093/pasj/psx062}, \href
  {https://ui.adsabs.harvard.edu/abs/2018PASJ...70S..29S} {70, S29}

\bibitem[\protect\citeauthoryear{Storkey, Hambly, Williams  \& Mann}{Storkey
  et~al.}{2004}]{elliptical_hough_transform}
Storkey A.~J.,  Hambly N.~C.,  Williams C. K.~I.,   Mann R.~G.,  2004, \mn@doi
  [Monthly Notices of the Royal Astronomical Society]
  {10.1111/j.1365-2966.2004.07211.x}, 347, 36

\bibitem[\protect\citeauthoryear{{The Dark Energy Survey Collaboration}}{{The
  Dark Energy Survey Collaboration}}{2005}]{DES}
{The Dark Energy Survey Collaboration} 2005, The Dark Energy Survey (\mn@eprint
  {arXiv} {astro-ph/0510346})

\bibitem[\protect\citeauthoryear{Vapnik}{Vapnik}{1979}]{svm}
Vapnik V.,  1979, Reconstruction of Dependences from Empirical Data

\bibitem[\protect\citeauthoryear{{de Jong}, {Verdoes Kleijn}, {Kuijken}  \&
  {Valentijn}}{{de Jong} et~al.}{2013}]{kilo-degree_survey}
{de Jong} J. T.~A.,  {Verdoes Kleijn} G.~A.,  {Kuijken} K.~H.,   {Valentijn}
  E.~A.,  2013, \mn@doi [Experimental Astronomy] {10.1007/s10686-012-9306-1},
  \href {https://ui.adsabs.harvard.edu/abs/2013ExA....35...25D} {35, 25}

\makeatother
\end{thebibliography}




\appendix

\section{Overview of Methods}

\subsection{CAS Swinburne}
\label{sec:cas_swinburne_appendix}

\begin{figure*}
	\includegraphics[width=\textwidth]{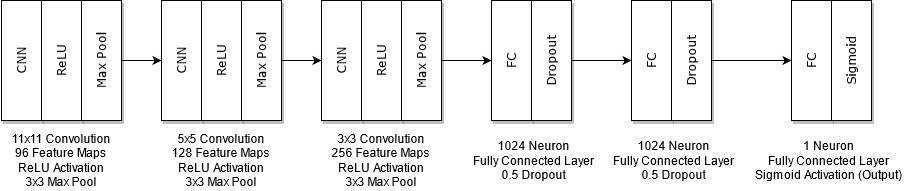}
    \caption{This is a graphical representation of the `CAS Swinburne' model described in Section~\ref{sec:cas_swinburne} and Appendix~\ref{sec:cas_swinburne_appendix}. Reproduced from \citet{cas_swinburne}.}
    \label{fig:cas_swinburne}
\end{figure*}

This model was based on AlexNet \citep[][]{alex_net}. The input image is first passed through three convolutional layers, with kernel sizes of 11, 5, and 3 respectively and 96, 128, and 256 feature maps respectively. Each convolutional layer was followed by a \gls{ReLU} activation function and a 3x3 max pooling layer. The output from the last max pool was put into two successive fully-connected layers, with 1,024 neurons each. A dropout layer with 0.5 probability was added after each fully-connected layer \citep[][]{cas_swinburne}. This architecture is shown graphically in Fig.~\ref{fig:cas_swinburne}.

\subsection{LASTRO EPFL}
\label{sec:lastro_epfl_appendix}

\begin{figure*}
	\includegraphics[width=\textwidth]{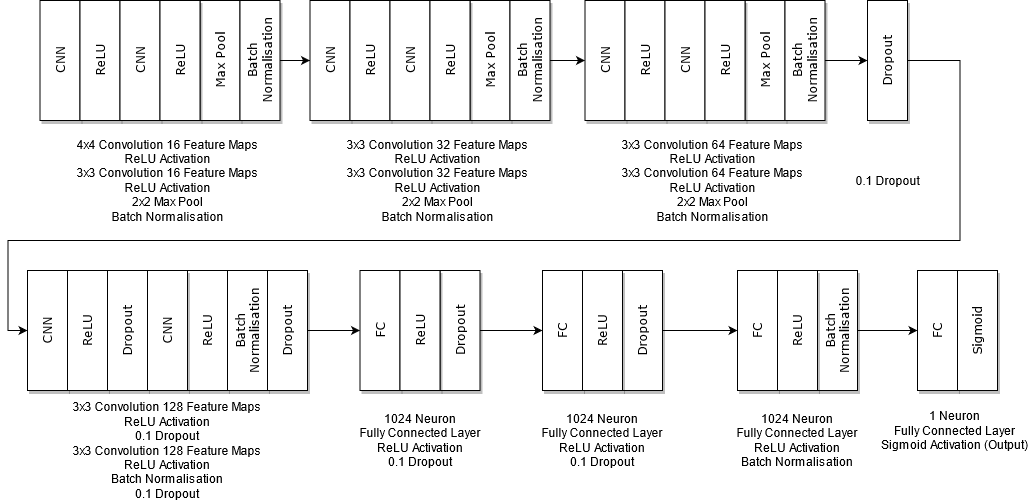}
    \caption{This is a graphical representation of the `lastro\_epfl' model described in Section~\ref{sec:lastro_epfl} and Appendix~\ref{sec:lastro_epfl_appendix}. Reproduced from \citet{lastro_epfl}.}
    \label{fig:lastro_epfl}
\end{figure*}

This model follows a somewhat similar architecture to that described in Section~\ref{sec:cas_swinburne}, however has significantly more layers. All layers in this model use a \gls{ReLU} activation, unless specified otherwise.
This model starts off with 3 blocks, where each block consists of two consecutive convolutional layers, followed by a max pooling layer and a batch normalisation layer.
The first block's convolutional layers use a kernel size of 4 and 3 respectively, with 16 features each. The convolutional layers in the second and third blocks all use a kernel size of 3, with the second block having 32 features, and the third having 64. As specified, all three blocks are followed by a max pooling and a batch normalisation layer.
After the third block, a dropout layer is added to reduce the possibility of overfitting.
A convolutional layer with a kernel size of 3 and 128 features is added, followed by another dropout layer. This is followed by another convolutional layer of the same specifications, this time followed by a batch normalisation layer and another dropout layer.\
The last layer's output is flattened and passed to a triple of fully-connected layers, with a dropout layer between each fully-connected layer. Batch normalisation is added after the last fully-connected layer.
The model's output is obtained by passing the output of the last batch normalisation to a fully-connected layer, with a single neuron, with a sigmoid activation function \citep[][]{lastro_epfl}. This architecture is shown graphically in Fig.~\ref{fig:lastro_epfl}.

\subsection{CMU DeepLens}
\label{sec:cmu_deeplens_appendix}

\begin{figure*}
	\includegraphics[width=\textwidth]{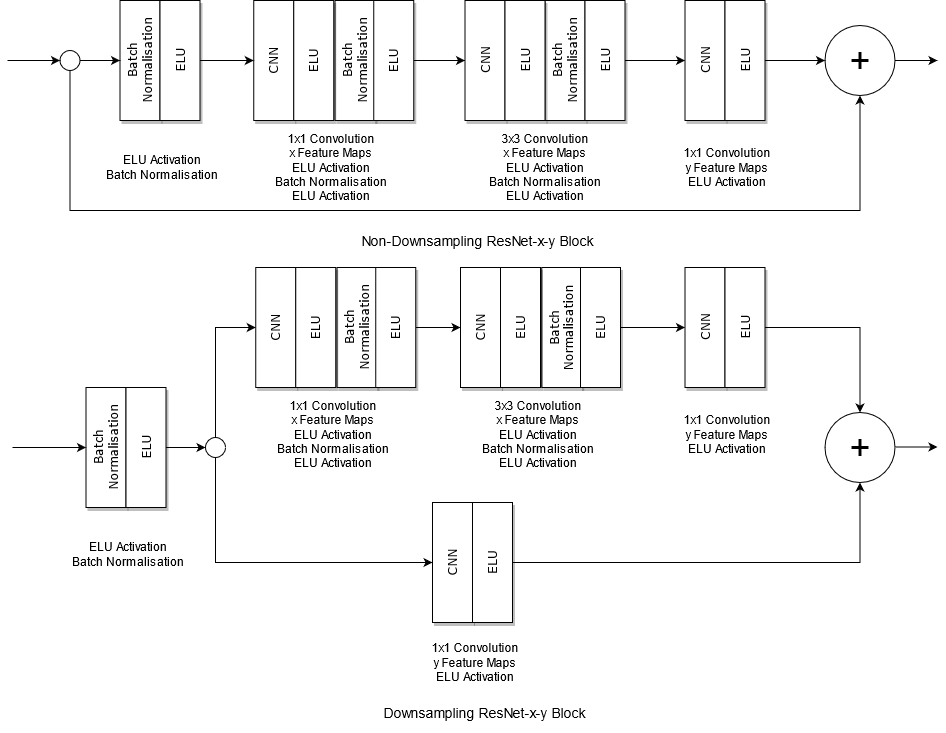}
    \caption{This is a graphical representation of the two types of `ResNet blocks' used by the `CMU DeepLens' model described in Section~\ref{sec:cmu_deeplens} and Appendix~\ref{sec:cmu_deeplens_appendix}. Reproduced from \citet{cmu_deeplens}.}
    \label{fig:cmu_deeplens_resnet_blocks}
\end{figure*}

As shown in Fig.~\ref{fig:cmu_deeplens_resnet_blocks}, two different types of `ResNet blocks' are used by this model, one which keeps the original resolution of the image, and another which downsamples the image by a factor of 2. In the case where the image is not downsampled, a copy of the input to the ResNet block is stored. The input is also passed through the triple of Batch Normalisation, Non Linearity (\gls{ELU}), and a Convolutional Layer three times. The result of these 9 layers is summed with the original input to the ResNet block, and returned as the output. In the case where downsampling is employed, the input to the ResNet block first goes through Batch Normalisation and Non Linearity (\gls{ELU}), after which a copy of the current tensor is stored for later use. This is followed by a Convolutional Layer with a stride of 2, and another two `Batch Normalisation, \gls{ELU} and Convolutional Layer' triples. The output of the last convolutional layer is summed with the aforementioned copy of the tensor at an earlier stage, after it has gone through a convolutional layer with stride 2, and returned as the block's output.

\begin{figure*}
	\includegraphics[width=\textwidth]{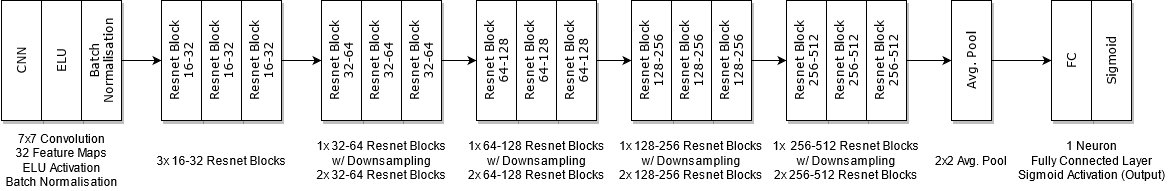}
    \caption{This is a graphical representation of the `CMU DeepLens' model described in Section~\ref{sec:cmu_deeplens} and Appendix~\ref{sec:cmu_deeplens_appendix}. Reproduced from \citet{cmu_deeplens}.}
    \label{fig:cmu_deeplens}
\end{figure*}

The CMU DeepLens model is structured as follows. The Image is first passed through a convolutional layer with a  kernel size of 7, with 32 features, using an \gls{ELU} activation function, and is followed by a Batch Normalisation Layer.
This is then followed by 3 ResNet blocks, each with 32 features. This is followed by another 4 sets of `3 ResNet blocks'. Each of these sets starts with a downsampling ResNet block, followed by 2 `regular' ResNet blocks. The features of each ResNet block in each set are 64, 128, 256 and 512 respectively.
The output from the last ResNet block is passed through an Average Pooling layer, and the model's prediction is finally computed by a fully connected layer with one neuron and a sigmoid activation \citep[][]{cmu_deeplens}. This architecture is shown in Fig.~\ref{fig:cmu_deeplens}.

\subsection{WSI-Net}
\label{sec:wsi_net_appendix}

\begin{figure*}
	\includegraphics[width=\textwidth]{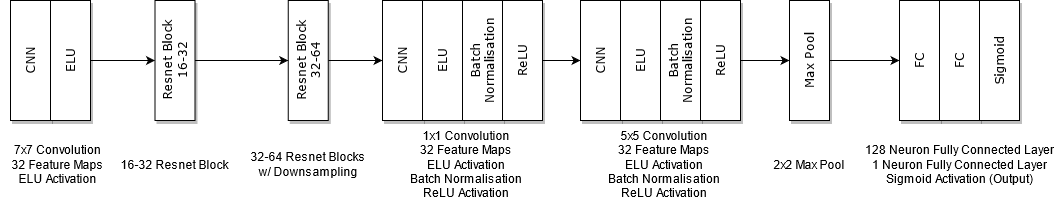}
    \caption{This is a graphical representation of the `WSI-Net' model described in Section~\ref{sec:wsi_net} and Appendix~\ref{sec:wsi_net_appendix}. Reproduced from \citet{wsi_net}.}
    \label{fig:wsi_net}
\end{figure*}

The first layer is a Convolutional Layer with a kernel size of 7, 32 features and an \gls{ELU} activation.
This is then followed by two ResNet blocks of 32 and 64 features respectively. These ResNet blocks used are the same as those described in Section~\ref{sec:cmu_deeplens}.
Following the two ResNet blocks is another Convolutional Layer with a kernel size of 1, 32 features and an \gls{ELU} activation.
This is followed by a Batch Normalisation Layer, and a \gls{ReLU} activation.
A Convolutional Layer with kernel size 5, 32 features and an \gls{ELU} activation is used next, again followed by a Batch Normalisation Layer as well as a \gls{ReLU} activation.
A Max Pooling Layer is added on next, followed by a Fully Connected Layer with 256 neurons.
The final classification is obtained by another Fully Connected Layer with 1 neuron, and a sigmoid activation \citep{wsi_net}. This architecture is shown in Fig.~\ref{fig:wsi_net}.

\subsection{LensFlow}
\label{sec:lens_flow_appendix}

\begin{figure*}
	\includegraphics[width=\textwidth]{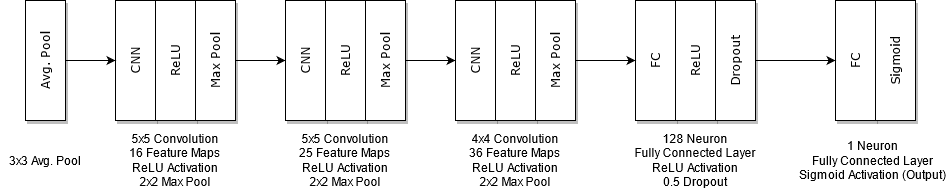}
    \caption{This is a graphical representation of the `LensFlow' model described in Section~\ref{sec:lens_flow} and Appendix~\ref{sec:lens_flow_appendix}. Reproduced from \citet{lens_flow}.}
    \label{fig:lens_flow}
\end{figure*}

In this model, the first operation carried out on the input image is an Average Pool with a pool size of 3x3 and a stride of 3.
This is followed by a Convolutional Layer with a kernel size of 5 and 16 features, and a max pooling layer with a pool size of 2 and a stride of 2.
This is again followed with another two `Convolutional Layer + Max Pool' pairs, where the convolutional layers have a kernel size of 5 and 25 features, and a kernel size of 4 and 36 features, respectively, and both max pools have a pool size of 2 and a stride of 2.
The last max pool layer is fed into a Fully Connected layer with 128 neurons and a \gls{ReLU} activation. During training, this layer is followed by a dropout layer with 0.5 probability.
The final output is obtained from a Fully Connected layer with 1 neuron, and a sigmoid activation \citep[][]{lens_flow}. This architecture is shown in Fig.~\ref{fig:lens_flow}.

\subsection{Lens Finder}
\label{sec:lens_finder_appendix}

\begin{figure*}
	\includegraphics[width=0.8\textwidth]{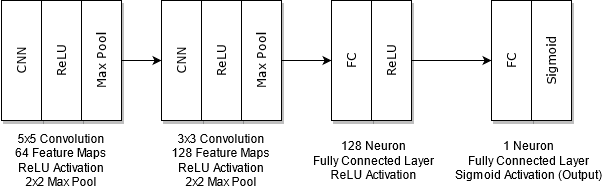}
    \caption{This is a graphical representation of the `LensFinder' model described in Section~\ref{sec:lens_finder} and Appendix~\ref{sec:lens_finder_appendix}. Reproduced from \citet{lens_finder}.}
    \label{fig:lens_finder}
\end{figure*}

The LensFinder model has a relatively simplistic architecture, when compared to some of the solutions presented in this paper, however holds its weight with the score it obtains.
The paper doesn't state specific values for the hyper parameters of each layer in the model, the values presented here are what were found to work best, empirically.
The model starts with a convolutional layer, with a kernel size of 5 and 64 features. A \gls{ReLU} activation function is used. The result is fed into a Max Pooling layer. The output is then passed into another convolutional layer with a kernel size of 3, and 128 features. Here again, a \gls{ReLU} activation is used and the output goes into a Max Pooling layer.
This is connected to a Fully Connected layer with 128 neurons, and a \gls{ReLU} activation. This output is connected to the final Fully Connected layer.
In the original paper, a softmax activation is used, however since this is a binary classification problem, only 1 neuron is used in this layer, and a sigmoid activation is used instead \citep{lens_finder}. This architecture is displayed in Fig.~\ref{fig:lens_finder}.

\bsp	
\label{lastpage}
\end{document}